\documentclass{ws-procs975x65}

\begin{document}

\title{Modelling Extreme-Mass-Ratio Inspirals using Pseudospectral Methods.}

\author{Priscilla Ca\~nizares and Carlos F.~Sopuerta}

\address{$^{1}$Institut de Ci\`encies de l'Espai (CSIC-IEEC), \\
Facultat de Ci\`encies, Campus UAB, Torre C5 parells, 
Bellaterra, 08193 Barcelona, Spain}

\begin{abstract}
We introduce a new time-domain method for computing the self-force acting on a scalar 
particle in a Schwarzschild geometry. The principal feature of our method consists in 
the division of the spatial domain into several subdomains and locating 
the particle at the interface betweem two them. In this way, we avoid the need of 
resolving a small length scale associated with the presence of a particle in the 
computational domain and, at the same time, we avoid numerical problems due to the 
low differentiability of solutions of equations with point-like singular behaviour.
\end{abstract}

\keywords{Gravitational Waves Astronomy; Extreme-Mass-Ratio Inspirals; Self-Force.}

\bodymatter

\section{Motivation} \label{Motivation}
One of the main sources of gravitational radiation for the future space-based 
gravitational-wave observatory LISA\cite{LISA} are the capture, and posterior inspiral, 
of stellar-mass compact objects (SCOs) into massive black holes (MBH) located at 
galactic centers. Since the masses of interest for the SCO are around $m = 1-10^2 M_{\odot}$, 
and for the MBH are in the range $M= 10^4-10^7 M_{\odot}$, the mass-ratio for these systems is  
$\mu=m/M \sim 10^{-7} -10^{-2}$. For this reason, they are called \emph{Extreme-Mass-Ratio 
Inspirals} (EMRIs). During the inspiral phase, an EMRI losses energy and angular momentum 
via the emission of gravitational waves (GWs).  LISA will be able to detect GW signals of 
$10-10^3$ EMRI$/yr$ up to distances with $z \lesssim 1$\cite{AmaroSeoane:2007aw}.  
These signals will be hidden in the LISA instrumental noise and in the GW foreground 
produced mainly by compact binaries in the LISA band.  Thus, in order 
to extract the EMRI signals we need a very accurate theoretical knowledge of the gravitational 
waveforms.  The main difficulty in producing those waveforms is the description of the 
gravitational effects of the SCO on its own trajectory.  These effects produce deviations
in the motion of SCO, which is not longer a geodesic around the MBH, which can be pictured 
as the action of a local force, the \emph{self-force}. 
Here, we review the results of recent work\cite{Canizares:2008dp,Canizares:2009ay}~where 
a new time-domain technique for the computation of the self-force has been proposed.
 
\section{Top tips for an efficient time-domain self-force computation} \label{Toptips}
Due to the complexity of the gravitational EMRI problem\cite{Mino:1997nk,Quinn:1997am}  
we have developed our new techniques for self-forse computations using a simplified model
that contains all the ingredients of the gravitational case.  It consists of a charged scalar 
particle (the SCO) in a circular motion around a non-rotating MBH.  This provides a good 
test bed to test our techniques before extending them to the gravitational case.  

The equations of a (scalar, electromagnetic, gravitational) field on a Schwarzschild background
inherit the spherical symmetry of the geometry.  Therefore, we can decompose the field in 
harmonic modes, eliminating the angular dependence, so that each of them satisfies a decoupled
1+1 wave-type equation.  In contrast with the behavior of the full field, each harmonic mode
turns out to be finite at the particle location, which is very useful to regularize the field
mode by mode using the {\em mode-sum} regularization scheme\cite{Barack:1999wf,Barack:2001gx,Barack:2002mha}. 
Then, we can obtain a regular field by adding all the regularized harmonic modes, from which we
can compute the self-force acting on the particle.  It is then very important to develop efficient 
techniques to compute precisely the harmonic modes near or at the particle location so that we can 
estimate very precisely the self-force via the mode-sum scheme.  

Recently, we presented\cite{Canizares:2008dp,Canizares:2009ay} a new time-domain technique to 
solve the wave equations for the harmonic modes in an efficient and precise way.  It consists in
a multi-domain framework where the particle is always located at the interface between two 
subdomains.  This setup has two important advantages: (i) We do not need to resolve a small scale
associated with the presence of a point-like singularity inside the computational domain and 
(ii) to avoid the negative effects in the numerical computations of the low 
differentiability of the solution.   The main idea is that the wave equations in the different
domains are source-free and then do not see point-like singularities, which ensure good
differentiability properties and hence, good numerical convergence properties.  The way the
different domains are communicated is through analytical junction conditions dictated by the
field equations themselves\cite{Sopuerta:2005gz}. 

Regarding the numerical implementation, we perform the spatial discretization using a 
Chebyschev-Lobatto \emph{Pseudospectral} Collocation Method\cite{Boyd} (PCM).  We evolve a
first-order system of equations, obtained from the reduction of the wave-type equations,
that allow us to impose the junction conditions on the characteristic fields of this system.
In practice, this is imposed via the \emph{penalty method}\cite{Hesthaven:2000jh}, which 
drives the system dynamically to satisfy the junction conditions\cite{Canizares:2008dp,Canizares:2009ay}.
The convergence properties of the PCM are very sensitive to the smoothness properties of the
solution, which gives more importance to the multi-domain techniques we are using.  In this sense,
it has been shown\cite{Canizares:2008dp,Canizares:2009ay} that our methods are able to resolve 
with precision the field on the particle location with a reasonably low computational cost, showing
that indeed the method is well suited for time-domain computations of the self-force. 
Up to now, we have done calculations for circular orbits.  Comparing our results with others in 
the literature\cite{Haas:2006ne,DiazRivera:2004ik}, we have found that they agree with a high
degree of precision even when we use a relatively low number of collocation points.  For instance,
at the last stable circular orbit ($r^{}_p = 6 M$), our computation of the radial derivative of 
the field, the only one that requires regularization in the circular case, coincides with 
the values obtained in other time-domain and frequency-domain calculations with a relative error of 
the order of $0.2\%$.  Our calculations use between 12 and 24 subdomains (for the discretization of
the radial direction in terms of the tortoise coordinate) and 50 collocation points per domain.
The average time for a full self-force calculation (which involves the calculation of 231 harmonic
modes) in a computer with two Quad-Core Intel Xeon processors at 2.8 GHz is always in the range 
20-30 minutes\cite{Canizares:2008dp,Canizares:2009ay}.  These calculations can be further optimized
by distributing the subdomains and collocation points so that the resolution is adapted to the
physical problem.  The calculations can be easily parallelized, either by spreading the work
of the harmonic modes or that of the subdomains.

Looking at the future, we are currently extending these techniques to eccentric orbits.  This 
has required some modifications to keep the particle fixed at a node between subdomains, and
results of the calculations will be published elsewhere\cite{Canizares_Sopuerta_Jaramillo}.
The next step will be the gravitational case, where the challenge comes from the fact that each
harmonic mode is described by a set of coupled 1+1 wave type equations.

\section*{Acknowledgments}
PCM is supported by a predoctoral FPU fellowship of the Spanish Ministry of
Science and Innovation (MICINN).
CFS acknowledges support from the Ram\'on y Cajal Programme of the
Ministry of Education and Science of Spain and by a Marie Curie
International Reintegration Grant (MIRG-CT-2007-205005/PHY) within the
7th European Community Framework Programme.


\begin{thebibliography}{10}

\bibitem{LISA}
{ LISA}: {\tt http://www.esa.int}, {\tt http://lisa.jpl.nasa.gov}.

\bibitem{AmaroSeoane:2007aw}
P.~Amaro-Seoane {\em et~al.}, {\em Class. Quant. Grav.} {\bf 24}, R113 (2007).

\bibitem{Canizares:2008dp}
P.~Canizares and C.~F. Sopuerta, {\em J. Phys. Conf. Ser.} {\bf 154}, 012053
  (2009).

\bibitem{Canizares:2009ay}
P.~Canizares and C.~F. Sopuerta, {\em Phys. Rev.} {\bf D79}, 084020 (2009).

\bibitem{Mino:1997nk}
Y.~Mino, M.~Sasaki and T.~Tanaka, {\em Phys. Rev.} {\bf D55}, 3457 (1997).

\bibitem{Quinn:1997am}
T.~C. Quinn and R.~M. Wald, {\em Phys. Rev.} {\bf D56}, 3381 (1997).

\bibitem{Barack:1999wf}
L.~Barack and A.~Ori, {\em Phys. Rev.} {\bf D61}, 061502 (2000).

\bibitem{Barack:2001gx}
L.~Barack, Y.~Mino, H.~Nakano, A.~Ori and M.~Sasaki, {\em Phys. Rev. Lett.}
  {\bf 88}, 091101 (2002).

\bibitem{Barack:2002mha}
L.~Barack and A.~Ori, {\em Phys. Rev.} {\bf D66}, 084022 (2002).

\bibitem{Sopuerta:2005gz}
C.~F. Sopuerta and P.~Laguna, {\em Phys. Rev.} {\bf D73}, 044028 (2006).

\bibitem{Boyd}
J.~P. Boyd, {\em Chebyshev and Fourier Spectral Methods} (Dover, New
  York, 2001).

\bibitem{Hesthaven:2000jh}
J.~S. Hesthaven, {\em Appl. Numer. Math.} {\bf 33}, 23 (2000).

\bibitem{Haas:2006ne}
R.~Haas and E.~Poisson, {\em Phys. Rev.} {\bf D74}, 044009 (2006).

\bibitem{DiazRivera:2004ik}
L.~M. Diaz-Rivera, E.~Messaritaki, B.~F. Whiting and S.~Detweiler, {\em Phys.
  Rev.} {\bf D70}, 124018 (2004).

\bibitem{Canizares_Sopuerta_Jaramillo}
P.~Ca\~nizares, C.~F. Sopuerta and J.~L. Jaramillo, {\em in preparation}
  (2010).

\end{thebibliography}

\end{document}